\documentclass[aps,prb,twocolumn,floatfix]{revtex4-1}
\usepackage{graphicx,epsfig,array}
\begin{document}
\title{A Real Space Description of Magnetic Field Induced Melting in the
Charge Ordered Manganites:~II. the Disordered Case}

\author{Anamitra Mukherjee$^{1}$ and Pinaki Majumdar$^2$}
\affiliation{$^1$Department of Physics and Astronomy, University of British
Columbia, Vancouver, BC, Canada, V6T 1Z1}

\affiliation{$^2$Harish-Chandra  Research Institute,
 Chhatnag Road, Jhusi, Allahabad 211019, India}

\date{\today}

\begin{abstract}
We study the effect of A site disorder on magnetic field induced melting of charge order (CO) in half doped manganites using a Monte-Carlo technique. Strong A-site disorder destroys CO even without an applied field. At moderate disorder, the zero field CO state survives but has several intriguing features in its field response. Our spatially resolved results track the broadening of the field melting transition due to disorder and explain the unusual dependence of the melting scales on bandwidth and disorder. In combination with our companion paper on field melting of charge order  in clean systems we provide an unified understanding of CO melting across all half doped manganites.
\end{abstract}

\maketitle

\section{Introduction}

Strong correlation among spin, charge, orbital and lattice degrees of freedom has made manganites a very active area of experimental\cite{mang-book,tok-rev} and theoretical \cite{dagotto1,edmc,other_theory} research. The strong coupling among different degrees of freedom leads, not only to novel phases, but also to phase separation tendencies. The interplay of this phase separation tendency and disorder lies at the heart of the phenomenon of colossal magnetoresistance (CMR)\cite{dag-1-cmr,dag-2-cmr}. 

Metal insulator transition by varying temperature\cite{PhysRevLett.108.237202,tok_melt1,tok_melt2,tok_melt3}, electric\cite{PhysRevB.89.014425,PhysRevB.88.024415,current-2} or magnetic\cite{kuwahara-1,kuwahara-2,cheong,parisi-1,parisi-2,parisi-3,parisi-4} fields is of great interest not just for CMR effect but also for understanding how different participating degrees of freedom respond to time dependent perturbation. This naturally leads to questions on the relaxation of competing phases and nonequilibrium response. Disorder is unavoidable in these materials and add further complications by suppressing ordering tendencies and creating local pinning centers. These issues are particularly important in understanding pump probe and optical excitation experiments\cite{dyn-1, dyn-2, dyn-3,Nature_Materials_Polli:2007di, PhysRevB.80.115128} apart from thermal and magnetic field cycling induced phase transitions. In our companion paper\cite{clean-long} we have addressed issues of this nature for a clean system. This paper discusses disorder effects. 

Given the intense activity in the field of manganites in the past decade, many aspects such as colossal magnetoresistance, multiferroic properties, doping, temperature and disorder effects have been extensively studied. We refer to the companion paper\cite{clean-long} on the clean field melting problem for details of the existing literature in this field. Here we focus on half doped  manganites. 

The magnetic field induced melting of the charge order (CO) in the half doped manganites has been probed extensively. Experiments have mapped out hysteretic response and  bandwidth dependence of the `melting' field in systems with low \cite{respaud}, intermediate \cite{kuwahara-1,kuwahara-2,tok_melt1,tok_melt2,tok_melt3,other_exp} and strong \cite{akahoshi} disorder. The spatial nature has also been extensively investigated \cite{PhysRevLett.108.237202,chen-1,chen-2,trokiner,cheong,melt-exp2a,melt-exp2b,freitas,parisi}.

 Theoretical  results\cite{kp-am-pm1,satpathy,fratini,cep_hrk1} on clean system and our companion paper\cite{clean-long} set the stage for the present work on disorder effects. Our goal is to understand the strikingly different response of similar bandwidth charge ordered manganites to magnetic fields \cite{atfld1,atfld2,tomioka}, to capture the disorder induced rounding of melting transition and to explain the real space nature of the field melted state in presence of disorder.

We begin by quickly recapitulating the physics of the `clean' system. Bandwidth (BW), in the manganites, is controlled by the mean ionic radius of rare earth (A) and alkaline  earth (A') elements in A$_{0.5}$A'$_{0.5}$MnO$_3$. At low BW, half-doped manganites are insulators with in-plane checkerboard CO, $d_{x^2-r^2}/d_{y^2-r^2}$ orbital order (OO) (on sites with larger charge disproportionation), and CE type magnetic order. We will simply call this the CE-CO-I phase. Large BW materials, have a ferromagnetic metallic (FM-M) ground state. These two states, and an intermediate metallic A-type antiferromagnetic state, compete with each other at $x=0.5$ as BW is varied by A site substitution maintaining half filling.

Disorder in these materials are characterized by the size mismatch of the A and A'  elements. In this paper we will consider three families of manganites of the form Ln$_{0.5}$A'$_{0.5}$MnO$_3$ with A'= Ca, Sr and Ba, while Ln=La, Pr, Nd, Sm, Eu, etc.
Among these the 'Ca- family' has minimum disorder, followed by the Sr and the Ba families. 

In the low disorder Ca family, which we will consider as 'clean', the CE-CO-I state can be melted to a FM-M by a magnetic field. The BW dependence of this process, the nature of the transition and the real space properties of the field melted state has been described in detail in our companion paper. 
In the clean problem the magnetic field needed to melt the CE-CO-I state depends on the closeness of the zero field CE-CO-I and the FM-M. This dependence is controlled by the BW. In the Ca family the melting fields vary inversely proportional to the BW because larger BW weakens the CE-CO-I and vise versa. However this simple picture of BW control of CO melting fields break down in the 'Sr family'.  The BW for the Sr family is in the same range as for the Ca family (for the Ln's listed above), however the melting fields decrease with reducing BW. We show in this paper that disorder plays a key role in determining this response to magnetic fields. 

By studying the magnetic field  induced melting of the CE-CO-I in presence of different strengths of disorder we present the following results:

(i) We show that for small disorder, the field needed for melting the charge order increases with reducing BW in the experimentally relevant range of BW, a behavior is representative of the weakly disordered Ca family. For a range of intermediate disorder values, the magnetic field needed to melt the charge order decreases with reducing the BW after a small initial increase. This captures the behavior of the Sr-family. For large disorder, the CO is destroyed even at zero field  as in the Ba family. 

(ii) We capture the disorder induced rounding of the melting transition and correlate it with the real space description of the melting phenomenon.

(iii) And finally establish that a complete understanding of the melting problem, at any disorder, can be achieved by studying the zero field 'bandwidth-disorder' phase diagram.

The paper is organized as follows. In Section~II  we summarize the key
experimental results. In section~III we define the model and
describe the method of solution. In Section~IV A. we discuss the properties of the zero field disordered reference state and in Section~IV B $\&$ C present the effects of disorder on field melting phenomenon. We conclude in Section~V. 
\section{Experimental Results}
We summarize the main results on field induced melting on the CE-CO-I state and the effect of A site disorder on them. Doping the A site with A' of different radius creates an alloy of A and A'. This changes the mean Mn-O-Mn bond angle and hence the mean hopping integrals for the electrons. It also introduces spatial fluctuations in the same due to the alloying. Further the difference in the valence states of A and A' (e.g. A=Ca and A'=Eu) creates Coulomb scattering centers for the electrons. The variance of the distribution of A and A' quantifies the extent of disorder. For the Ca, Sr, and Ba families \cite{atfld1,atfld2,tomioka} with A'=La, Pr, Nd, Sm, Eu, etc., the variances are $\sim 10^{-3}\AA^2 $, $\sim 10^{-2}\AA^2 $ and $\sim10^{-1}\AA^2 $ respectively, close to half filling. Thus, the Ca family has low disorder and the Sr family is moderately disordered but the CE-CO-I state survives the disordering. Disorder in the Ba family is strong enough to destabilize the CE-CO-I state at zero fields\cite{akahoshi}. 
\begin{figure}[t]
	\vspace{.2cm}
	\centerline{
		\includegraphics[width=8.0cm,height=5.0cm,angle=0,clip=true]{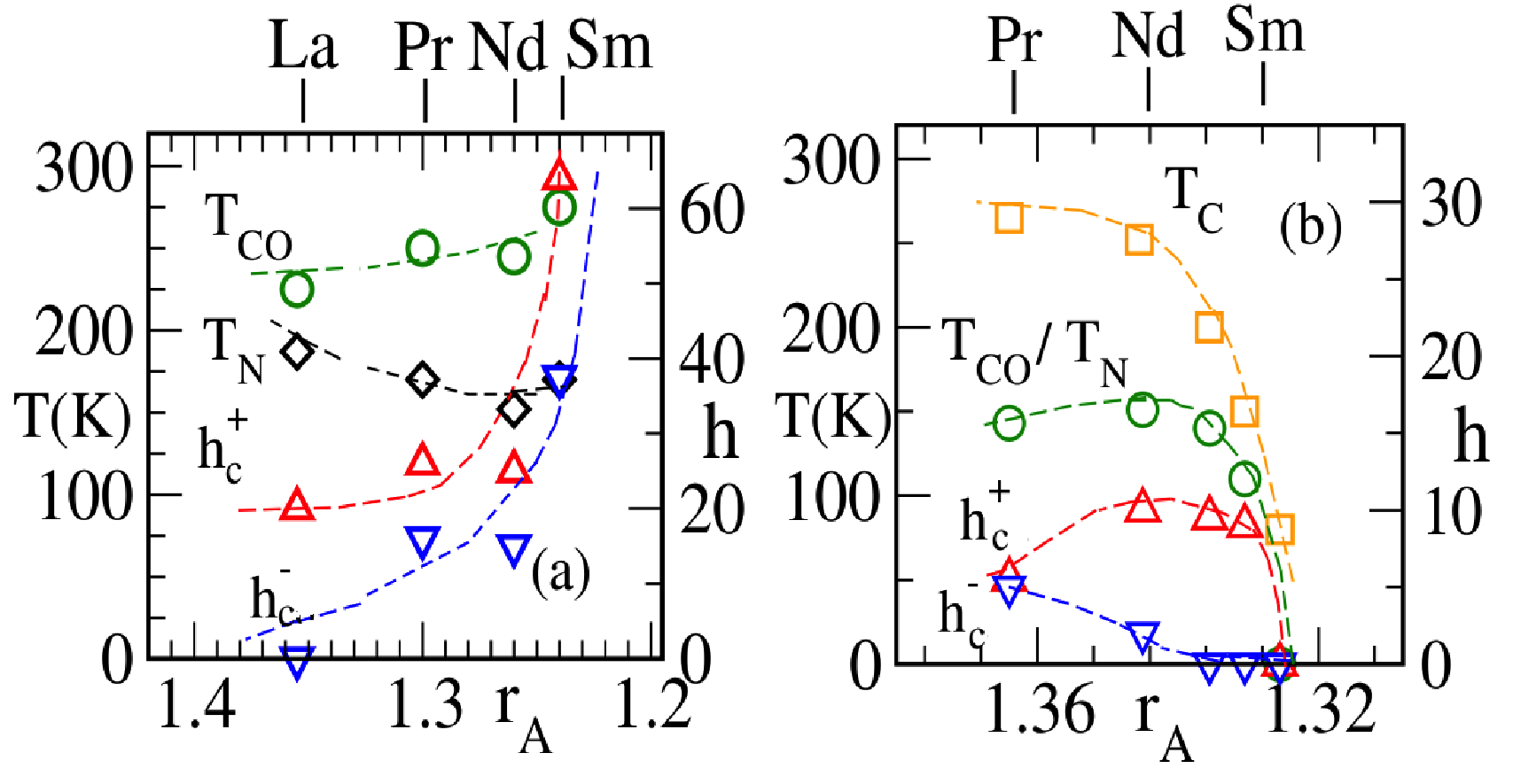}}
	\vspace{.2cm}
	\caption{Color online: The dependence of the zero field CO melting
		temperatures ($T_{CO}$), the (CE) Neel temperatures ($T_N$) and the magnetic fields needed to melt the CO at low temperature ($h_{c}^{\pm}$) on $ r_A$. (a) shows the data for the Ca family (Ln$_{0.5}$Ca$_{0.5}$MnO$_3$) and (b) for the Sr family (Ln$_{0.5}$Sr$_{0.5}$MnO$_3$).  The $Ln$ labels are indicated on the top and the corresponding $Ln$ radii ($r_A$) is indicated at the bottom. These plots are reconstructed from experimental data \cite{respaud,tok_melt1,tok_melt2,tok_melt3}. Decreasing $r_A$ implies reducing BW in all families.}
	\label{f-1}
\end{figure}
Fig. \ref{f-1} (a)-(b) show the key differences in the BW (or  $r_A$) dependence of the response to magnetic fields between the low disorder  (Ca) and the moderately disordered (Sr) families. Here, decreasing the Ln radius, $r_A$, implies reducing BW.  These plots, reconstructed from experimental data \cite{respaud,tok_melt1,tok_melt2,tok_melt3} show as a function of $r_A$, the evolution of magnetic fields $h^{+}_{c} $($h^{-}_{c} $) needed to melt (recover) the CO state while increasing (decreasing)  the field at low temperature. It also shows the corresponding zero field charge ordering (circles) and CE ordering (diamonds) temperatures as a function of $r_A$. (a) shows the data for the Ca family and (b) for the Sr family. As defined in the companion paper\cite{clean-long}, the CO  melting fields are defined as the magnetic field values at which the CO volume fraction changes by a large amount abruptly. There is a concomitant sharp drop in the resistivity ($\rho$) and also formation of a FM magnetic state. 

The labels on top show the A site dopant and (a) is for the Ln$_{0.5}$Ca$_{0.5}$MnO$_3$ family and (b) is for the Ln$_{0.5}$Sr$_{0.5}$MnO$_3$ family. Decrease in $r_A$ or reduction in BW  for the low disordered system Fig. \ref{f-1} (a) increases the CO melting temperatures and the switching fields $h_{C}^{\pm}$. For the range of $r_A$, 1.36-1.32$\AA$, while there is an increase in $T_{CO}$ and $h_{C}^{\pm}$ for the Ca family, the corresponding quantities in the Sr case initially rise a bit and then are very strongly suppressed and drops to zero at $r_A\sim 1.32$, which corresponds to Sm$_{0.5}$Sr$_{0.5}$MnO$_3$\cite{tok_melt1,tok_melt2,tok_melt3}. For materials with larger disorder such as the Ba family, the CE-CO-I state does not exist even at $h/t$=0  
\begin{figure}[t]
	\vspace{.2cm}
	\centerline{
		\includegraphics[width=9.0cm,height=8.0cm,angle=0,clip=true]{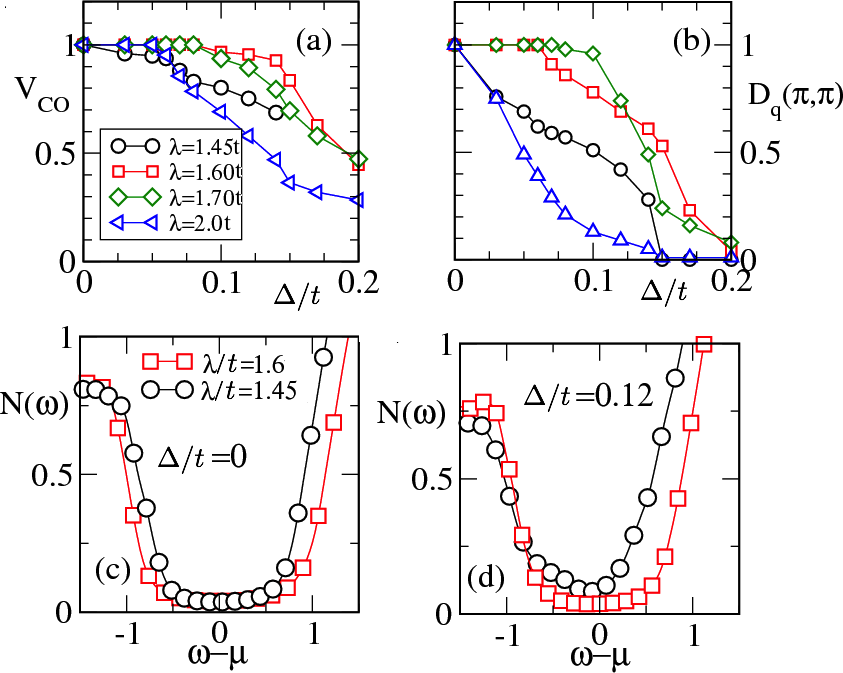}}
	\vspace{.2cm}
	\caption{Color online: The variation of $V_{CO}$ (a) and $D_{q}(\pi,\pi)$ (b), with disorder strength at various $\lambda$ values. The $V_{CO}$ and $D_q(\pi,\pi)$ fall with disorder most rapidly for the low (1.45) and large (2.0) $\lambda$, with the $D_{q}(\pi,\pi)$  being suppressed more quickly with disorder than $V_{CO}$. In contrast both $V_{CO}$ and $D_q(\pi,\pi)$ remain relatively robust up to $\Delta=0.1$ for $\lambda/t=1.6$ and $1.7$. We also show the density of states for $\lambda/t=1.45$ and $\lambda/t=1.6$ for clean (c) and disordered (d) cases. For $\Delta/t=0.12$, disorder changes the charge gap for $\lambda/t=1.45$ to a pseudogap. In contrast, the clear charge gap in the DOS for $\lambda/t=1.6$ survives at this disorder. The data is for $T/t=0.02$.}
	\label{f-2}
\end{figure}

Finally, field cycling experiments have found step like structures in the magnetization as a function of magnetic field in AF-I to FM-M transitions close to half doping\cite{step-5,step-4,step-2,step-3}. These step like features occur for low field cycling rates $\sim $1T/s that goes over to an abrupt transition for fast field sweep rates\cite{step-5} of $\sim 10^3$T/s . The step like features however depend on the sample quality indicating that the local competition between the FM-M and the AF-I that leads to the step like feature is affected by disorder and strain in the samples\cite{step-3}.

\begin{figure*}[t]
\vspace{.2cm}
\centerline{
\includegraphics[width=17.0cm,height=7.0cm,angle=0,clip=true]{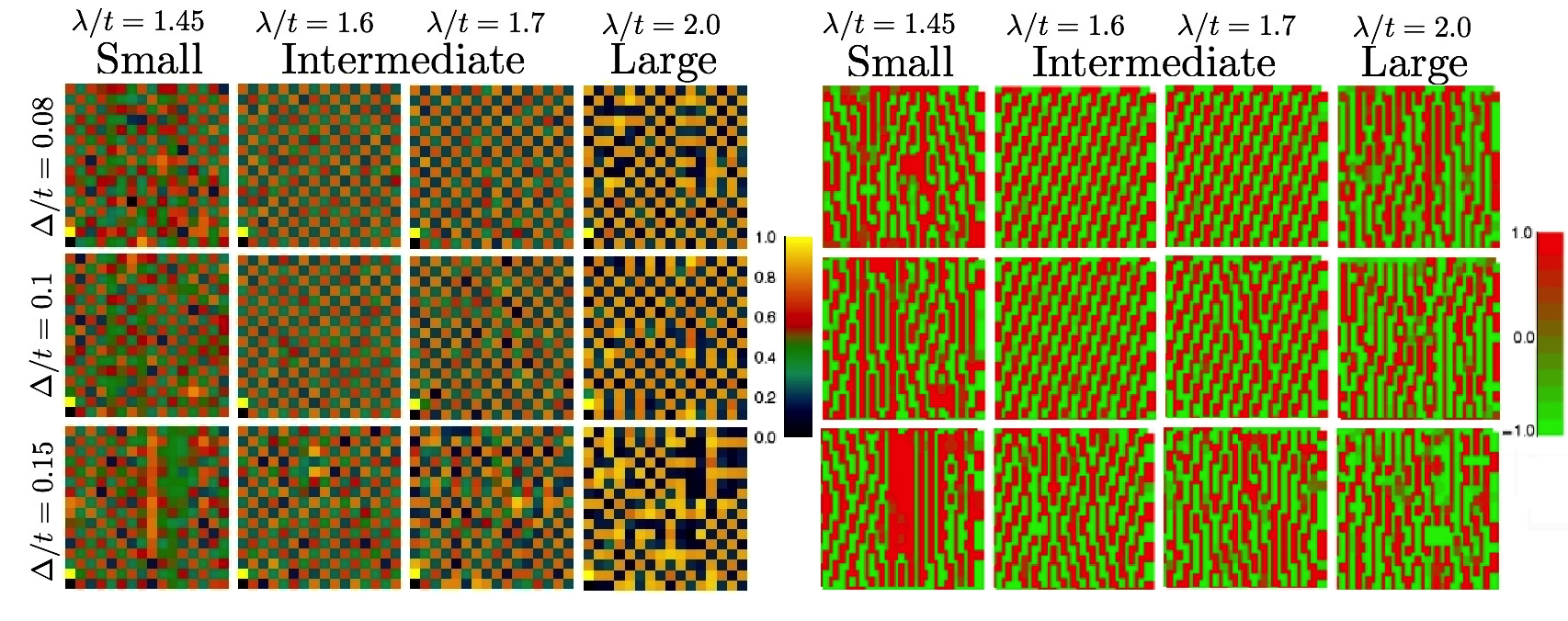} }
\vspace{.2cm}
\caption{Color online: (a) The spatial snapshots of charge density for four $\lambda/t$ values (1.45, 1.6, 1.7, 2.0), from left to right, at $\Delta/t$= 0.08, 0.10 and  0.15, from top to bottom. Yellow corresponds to $n_i=1$ at a site i and black implies $n_i=0$. (b) The real space snapshots of the magnetic state corresponding to the charge density profile shown in (a). In (b) minus one (green) indicates perfect antiferromagnetic bonds and plus one(red) indicates perfect ferromagnetic bonds. The $\lambda/t$ and $\Delta/t$ values and the $\lambda/t$ regime they belong to, are indicated in the figure. }
\label{f-3}
\end{figure*}

\section{Model and method}
We consider a two band model for $e_g$ electrons, Hund’s coupled to $t_{2g}$ derived core spins, in a two dimensional square lattice. The electrons are also coupled to Jahn-Teller phonons, while the core spins have an AF superexchange coupling between them. These ingredients are all necessary to obtain a CE-CO-I phase\cite{dag-ce,brink-1-half-dop-phases,dag-1-cmr,dag-2-cmr,dag-3-cmr}.
Similar approaches have been used in the past to study key aspects of the manganites\cite{brink-2-novel-phases,brink-1-mulf,brink-3-mulf,dag-0-mulf,dag-2-mulf,dag-3-mulf,brey-1-phases,brey-2-phases}. We include the effect of disorder through an on site potential.

\begin{eqnarray}
H &=& -\sum_{\langle ij \rangle \sigma}^{\alpha \beta}
t_{\alpha \beta}^{ij}
 c^{\dagger}_{i \alpha \sigma} c^{~}_{j \beta \sigma} 
+ \sum_i (\epsilon_i -\mu)n_i 
~ - J_H\sum_i {\bf S}_i.{\mbox {\boldmath $\sigma$}}_i \cr
&&
+ J\sum_{\langle ij \rangle}
{\bf S}_i.{\bf S}_j
 - \lambda \sum_i {\bf Q}_i.{\mbox {\boldmath $\tau$}}_i
+ {K \over 2} \sum_i {\bf Q}_i^2 
-h\sum_{i}{\bf S}_{i.z}
\nonumber
\end{eqnarray}
\noindent
Here, $c$ and $c^{\dagger}$ are annihilation and creation operators for $e_g$ electrons. $\alpha$ and $\beta $ are the two Mn $e_g$ orbitals $d_{x^2-y^2}$ and $d_{3z^2-r^2}$ respectively. $t_{\alpha\beta}^{ij}$ are the hopping amplitudes between nearest-neighbor Mn sites with the symmetry dictated form:~$t_{\alpha \alpha}^x= t_{\alpha \alpha}^y \equiv t$,~$t_{\beta \beta}^x= t_{\beta \beta}^y  \equiv t/3 $,~$t_{\alpha  \beta}^x= t_{\beta \alpha}^x \equiv -t/\sqrt{3} $,~$t_{\alpha \beta}^y= t_{\beta \alpha}^y  \equiv t/\sqrt{3} $, where $x$ and $y$ are spatial directions. The $e_g$ electron at a site i, ${\sigma}^{\mu}_i= \sum_{\sigma \sigma'}^{\alpha} c^{\dagger}_{i\alpha \sigma} \Gamma^{\mu}_{\sigma \sigma'} c_{i \alpha \sigma'}$, couples to the onsite core spin ${\bf S}_i$ via Hund's coupling J$_H$. Here the $\Gamma$'s are Pauli matrices. We  assume $J_H/t \gg 1$. $J$ is the superexchange interaction strength between the t$_{2g}$ spins. $\lambda$ is the coupling between the JT distortion ${\bf Q}_i = (Q_{ix}, Q_{iz})$ and the orbital pseudospin ${\tau}^{\mu}_i = \sum^{\alpha \beta}_{\sigma} c^{\dagger}_{i\alpha \sigma} \Gamma^{\mu}_{\alpha \beta} c_{i\beta \sigma}$, and $K$ is the lattice stiffness. We set $t=1$, $K=1$, and treat the ${\bf Q}_i$ and ${\bf S}_i$ as classical variables. The chemical potential $\mu$ is adjusted so that the electron density is held fixed to half filling.

We consider a lattice of Mn ions and treat the alloy disorder due to cationic substitution as a random potential $\epsilon_i$ at the Mn site picked from the distribution $P_A(\epsilon_i) = {1 \over 2}(\delta(\epsilon_i - \Delta) +  \delta(\epsilon_i + \Delta))$. The strength of disorder is quantified by the variance $\Delta$. 

We employ a real space exact diagonalization (ED) based Monte Carlo (MC) technique\cite{tca}  that allows us to perform calculations on system sizes of up to $40^{2}$. Disorder results are averaged over 30 disorder realizations. For reasons discussed in the companion paper,  we present all results for $J/t=0.12$.

\section{Results}
The crucial difference between the Ca and Sr families, which have similar BW, is disorder. So our first task is to determine the response of the CE-CO-I state to disorder at zero magnetic field. Since the field induced melting of the CO state depends on its robustness at zero field, this study is crucial. 

\subsection{Zero field}
We will use the measure of CO volume fraction ($V_{CO}$), the CO structure factors $D_{Q}(\pi,\pi)$, spin structure factor $S_q$ and real space snapshots of charge density and spins to study how disorder affects the CE-CO-I state. These quantities are defined in the Appendix. The BW variation, which in experiments is achieved by using $Ln$ atoms of different radii ($r_A$), is mimicked in our calculation by varying $\lambda$. Since the dimensionless electron phonon coupling is $\lambda/t$ and we set $t=1$, increasing $\lambda$ amounts to BW reduction and vice versa. 

\begin{center}
\textit{1. Bulk indicators:}\end{center}

Fig.\ref{f-2}(a) and (b) show the disorder averaged $V_{CO}$ (a) and the corresponding CO structure factors $D_{Q}(\pi,\pi) $ (b),  in the ground state as a function of disorder strength for various $\lambda$ values. In out companion paper\cite{clean-long} we have shown that for $J/t=0.12$, the CE-CO-I state exists beyond $\lambda/t=1.4$. For our discussion, we divide the range of $\lambda$ values into small ($\lambda/t \sim 1.4-1.55$), intermediate ($\lambda/t \sim 1.6-1.7$) and large ($\lambda/t >1.7$) ranges where the clean system has CE-CO-I as the ground state.  Here we discuss the response of the CE-CO-I phases belonging to these three regimes:

(i) We see that  both for small $\lambda/t(\sim 1.45)$ and large $\lambda/t( \sim 2.0)$, the volume fraction of charge order decreases rapidly with disorder strength ($\Delta/t$). For intermediate $\lambda/t(\sim 1.6,~1.7)$, the volume fraction of the CO regions remains robust till $\Delta/t=0.12 $. 

(ii) The CO structure factors ($D_{Q}(\pi,\pi)$), for small and large $\lambda/t$,  diminish much more quickly than the corresponding $V_{CO}$. This indicates that up to these values, disorder primarily introduces domain walls in the CO state at small and large BW. At  intermediate $\lambda/t$ regime, $V_{CO}$ and $D_{Q}(\pi,\pi) $ remain robust up to $\Delta/t\sim 0.12$ beyond which it is gradually destroyed. For larger disorder ($\Delta/t\sim 0.2$) this the CO becomes short range and, as seen in (a), has about $40 \%$ CO volume fraction.

(iii) The charge gap for $\lambda/t=1.45$ in the density of states (DOS) for the clean system, shown in (c), transforms into a pseudo gap state for $\Delta/t=0.1$ in (d), signaling formation of metallic puddles. In contrast for intermediate and large cases, the DOS shows the similar charge gap as in the clean case with some rounding at the band edges due to disorder. 

The lack of CO and hard charge gap for large $\lambda$ indicates pining of electrons at random sites leading to a disordered polaronic insulator, thereby gaining energy of $E_{JT}\equiv -\lambda^2/2K$ per such localization. These conclusions are further corroborated in real space snapshots of the charge and the spin configurations.
\\\\
\begin{center}
\textit{2. Real space evolution:} \end{center}

Fig. \ref{f-3} (a) shows snapshots of charge density ($n_i$) at three values of disorder, $\Delta/t=0.08$, $\Delta/t=0.1$ and $\Delta/t=0.15$ from top to bottom and for four values of  $\lambda/t(=1.45, 1.6, 1.7 $ and $2.0)$, increasing from left to right. Fig. \ref{f-3} (b) shows the snapshots of the corresponding spin states.
All the parameter points are CE-CO-I at $\Delta/t=0$. 
For disorder $\sim$ 0.08 and $\lambda/t(=1.45)$ (the far left column, top row in (a)) small metallic patches and domain walls begin to develop along with local disruption of the CE chains as seen in the corresponding panel (b). These metallic FM regions grow with increasing disorder (second and third rows of the far left column). 
At $\lambda/t(=2.0)$ (the far right column), clear domain walls start forming for $\Delta/t=0.08$ and $0.1$. Finally at disorder of $0.15$, the large $\lambda$ state forms a highly disordered polaronic insulator with randomly pinned charges. The corresponding CE phases starts out by forming domain walls and finally (at $\Delta/t=0.15$) forms small regions with $(\pi,\pi)$ correlations in regions of small charge density, as superexchange would dictate. The snapshots for the intermediate $\lambda$ cases the middle two panels shows that the CE-CO-I state is destroyed only at $\Delta/t=0.15$. Although the CE state at $\lambda/t=1.7$ begins forming some domain walls at $\Delta/t=0.1$.
\begin{figure}
\vspace{.2cm}
\centerline{
\includegraphics[width=8.0cm,height=8.0cm,angle=0,clip=true]{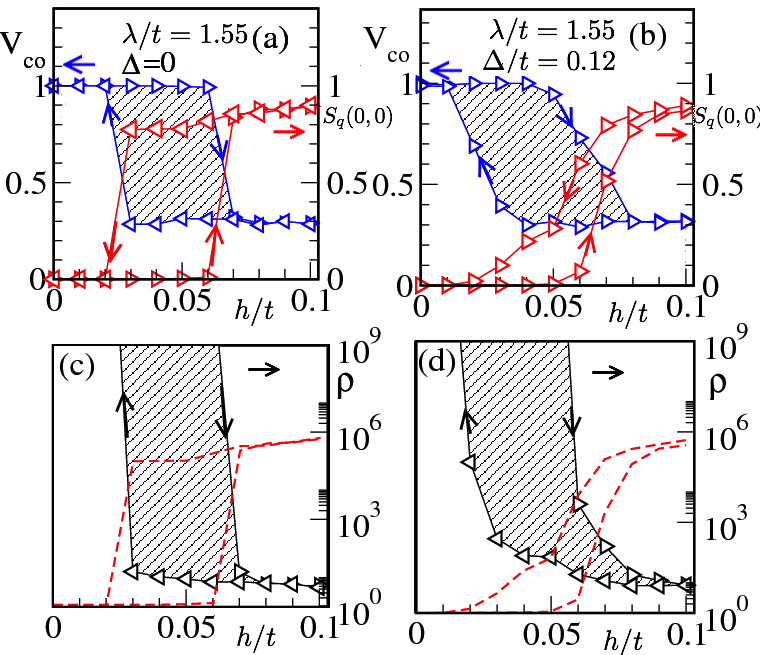} }
\vspace{.2cm}
\caption{Color online: The comparison of $V_{CO}$, $S_q(0,0)$ and resistivity ($\rho$) for the clean and the disordered cases for $\lambda/t=1.55$. The abrupt transition of $V_{CO}$ (blue curves) at $\Delta/t=0$ in (a),  is replaced by 'rounded' transitions in (b), the disordered case ($\Delta/t=0.12$) . The accompanying $S_q(0,0)$ (red curves) reaches close to saturation through a series of small steps in (b) as opposed to a sharp jump in (a). The resistivity (black curves) in (d) also shows similar behavior of changing in small steps in contrast to the sharp change in (c). For comparison of the switching locations we have overlaid the corresponding $S_q(0,0)$ in (c) and (d) with dashed line. The data is for $T/t=0.02$.}
\label{f-5}
\end{figure} 
However a general conclusion from the above analysis is that the CE-CO-I state is much more robust to disorder at intermediate $\lambda$ than for small and large $\lambda$ values. These systematics can be understood within a Landau like energy landscape proposed in the companion paper \cite{clean-long} on the clean field melting problem. There it was shown that the FM-M state has a wide domain of metastability and remains as a metastable state for the range of $\lambda$ values 1.45-1.55. Thus during thermal annealing for these $\lambda$ values, disorder traps the system partly in this FM-M state. At large $\lambda$, the CO state is inherently weak as the electrons are mostly  site-localized, with very small overlap between sites. This weak CO stiffness, that scales as ($t^4/E_{JT}^3$), is easily overcome even by weak disorder which disrupts the CO state. At intermediate $\lambda$, the CO is relatively robust because on the one hand this CE-CO-I state is energetically far away from the FM-M  and on the other, has greater CO stiffness than in the large $\lambda$ case.

\subsection{Field sweep in presence of disorder} 

\begin{figure}
\vspace{.2cm}
\centerline{
\includegraphics[width=9.0cm,height=3.0cm,angle=0,clip=true]{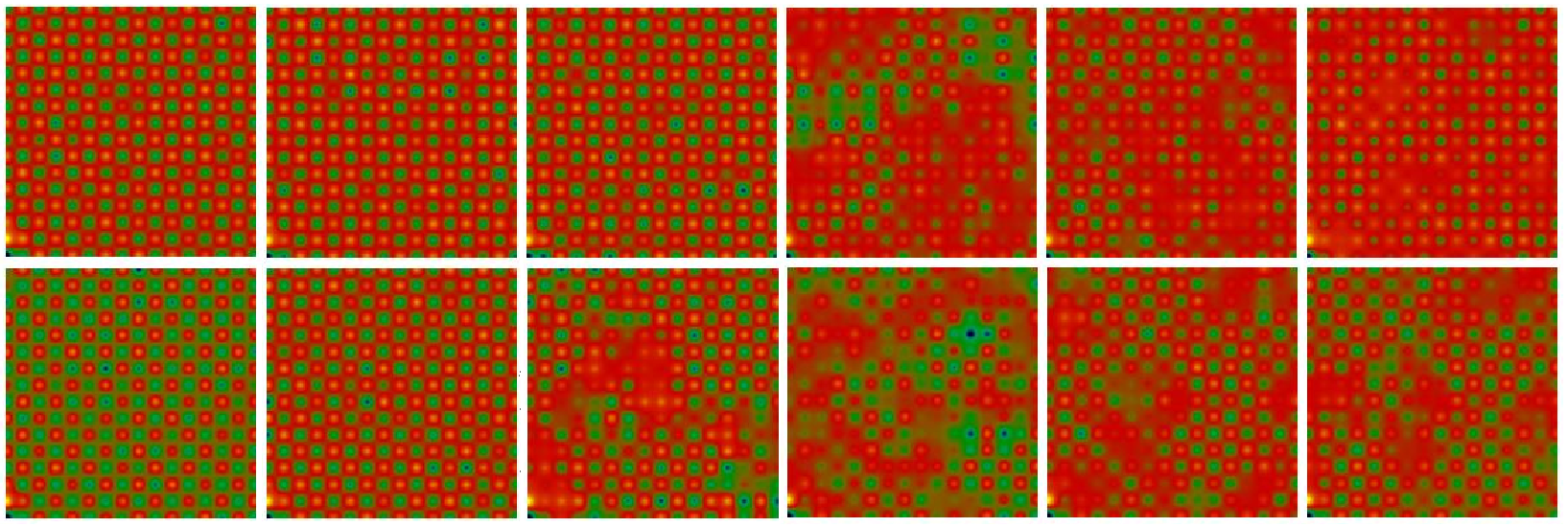} }
\vspace{.2cm}
\caption{Color online: Spatial snapshots of charge densities for $\lambda/t=1.55$ at $T/t=0.02$ for clean (top panel) and disordered ($\Delta/t=0.12$) (bottom panel) cases. The snapshots are shown for field values (h/t=0.04, 0.05, 0.06, 0.07, 0.08, 0.09) from left to right, for the increasing part of the field cycle. The corresponding $V_{CO}$ are shown in Fig. \ref{f-5} (a) and (b). In the clean case, the CO remains stable up to $h/t\sim 0.06$ and then collapses within a window of $\delta h/t=0.01$ to a state with about $30\%$ CO regions and $70\%$ FM-M. The disordered case however, starts losing CO at $h=0.05$ and collapses, within a window of $\delta h/t=0.04$ into a percolative metallic state with much large CO volume fractions that in the clean case.}
\label{f-6}
\end{figure} 
We now study the field response of the CE-CO-I state to magnetic field sweeps in presence of disorder. We track the magnetic field induced phase transition between the CE-CO-I and the FM-M as it takes place in the disordered background and contrast it with what we found in the clean case\cite{clean-long}.

Fig.\ref{f-5} and \ref{f-6} show this evolution of the zero field CE-CO-I state (for  $\lambda/t=1.55$) as we sweep the magnetic field on the zero field cooled system for $\Delta=0$ and $0.12$.  The CE-CO-I ground state is completely destroyed even at zero field for larger $\Delta/t\geq 0.15$.

Fig.\ref{f-5} measures the bulk properties $V_{CO}$, $S_q(0,0)$ and $\rho$, as a function of applied field in the clean (a), (c) and disordered (b), (d) cases. This gradual loss of  CO volume fraction in Fig.\ref{f-5} (b) and the abruptness of the transition in the clean case in Fig.\ref{f-5} (a) bring out the effect of disorder. The  resistivity in both cases switch to low values as the $V_{CO}$ switches. From the slope of the resistivity vs temperature curve (not shown),  we deduce that the low resistivity states are metallic.  Both  insulator to metal transition and the concomitant CE to FM transitions are gradual in the disordered case and show hysteresis similar to $V_{CO}$.  
Fig.\ref{f-6} shows the spatial profile of the charge density at different field values during the increasing part of a typical magnetic field sweep, $h$ increasing from left to right. 
The CO in Fig.\ref{f-6}, for the clean system (top panel) is seen to resist the field up to $h/t=0.06$, beyond which it abruptly goes to a percolative FM-M state. The disordered case (bottom panel) shows a more gradual trend in the melting, it starts by creating small metallic regions which grow over a window of field values to reach the final percolative metallic state.

\begin{figure}
\vspace{.2cm}
\centerline{
\includegraphics[width=9.0cm,height=5.0cm,angle=0,clip=true]{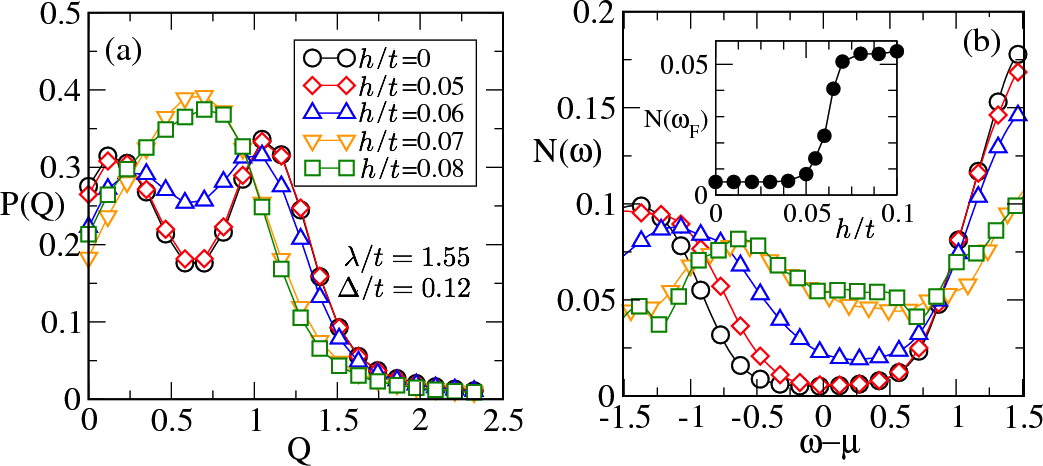} }
\vspace{.2cm}
\caption{Color online: (a) $P(Q)$ and (b) DOS at various magnetic field values for $\lambda/t=1.55$ and $\Delta/t=0.12$ at $T/t=0.02$. In (a) the $h/t=0$, two peak structure of P(Q) implied binary charge distribution that occurs in the CO state. With increasing field this gradually goes over to a broad distribution, signaling the melting of the CO state into a inhomogeneous state due to disorder. Inset in (b) shows the gradual increase in the weight at the Fermi level with increasing magnetic field.}
\label{f-7}
\end{figure}

Fig. \ref {f-7} shows the distribution of the lattice distortion P(Q) in (a) at different field values and $\Delta/t=0.12$ . Here the twin peak profile (the signature of the CO) at zero fields evolve into a broad hump, signaling the disorder induced coexistence. The well known transition in the clean case is abrupt, where the P(Q) peaks at Q= zero in the melted state. 
(b) shows DOS for $\Delta/t=0.12$ at different field values, where we see a gradual filling up of the charge gap with increasing field. The inset in (b) shows the weight at the Fermi energy as a function of the applied field which again depicts the gradual filling up of the charge gap instead of a abrupt transition expected in the clean case.

\subsection{Field melting systematics $\&$ the $\lambda-\Delta$ phase diagram}
We  now show that the behavior of the thermal and magnetic melting scales for the CO state with $\lambda$ or ($r_A$) for both the clean and the disordered cases, as seen in Fig.\ref{f-1}, can be understood in a single framework. 
 
\begin{figure}
 \vspace{.2cm}
 \centerline{
 \includegraphics[width=8.0cm,height=4.0cm,angle=0,clip=true]{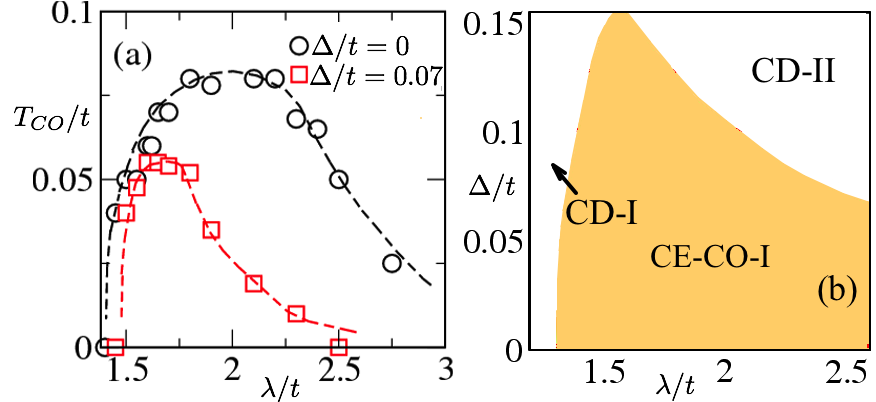} }
 \vspace{.2cm}
 \caption{Color online: (a) The variation of $T_{CO}$ with $\lambda$ at $\Delta/t=0$ and $\Delta/t=0.07$. The $T_{CO}$ shows a non monotonic behavior with $\lambda$. 
 The overall T$_{CO}$ scales is suppressed and the maxima shifts to a lower $\lambda$ value in the disordered case. (b) The $\lambda-\Delta$ phase diagram at $T/t=0.02$. The CE-CO-I region is sandwiched between charged disordered regions CD-1 and CD-2. These regions are defined in the text.}
 \label{f-4}
 \end{figure}

 Fig.\ref{f-4} (a) shows the $h/t=0$, T$_{CO}$ dependence on $\lambda$ for $\Delta/t=0$ and $\Delta/t=0.07$.  In the companion paper we have discussed that the CE-CO-I region, for J/t=0.12, is bounded by a metallic state at smaller $\lambda$ values. At large $\lambda$ the CE-CO-I stiffness weakens with increasing $\lambda$ as $t^4/E_{JT}^3$. The CE-CO-I state is most stable between these two limits, 
which explains the non monotonic dependence of  T$_{CO}$ on $\lambda$ at zero disorder. Disorder piles on this weakness of the charge order at small $\lambda$ (due to the presence metastable FM-M)  and at large $\lambda$ (due to weak CO stiffness) and suppresses the $T_{CO}$ scales at both ends as seen in  Fig.\ref{f-4} (a). With the disorder induced shrinking of the $\lambda$ range where CE-CO-I  can exists, we find that the peak of T$_{CO}$ too shifts to lower $\lambda$ values.
  
 From such data we construct the $\Delta-\lambda$ phase diagram, shown in  Fig.\ref{f-4} (b). The $\Delta/t=0$ charge ordered region occurs for  $\lambda\geq1.4$ which is the CE-CO-I / FM-M phase boundary in the clean problem\cite{clean-long}. 
 Disorder pushes this to higher $\lambda$. We denote the low $\lambda$ region without charge order  as CD-1.  On the large $\lambda$ side, the CO stiffness gets progressively weak with $\lambda$ making it susceptible to disorder. Thus at any finite disorder there is a critical $\lambda$, dictated by the competition of the pinning effects of disorder and the CO stiffness,  beyond which the CO state is destroyed in favor of a disordered polaronic insulator, CD-2. This critical $\lambda$ decreases with increasing disorder.  Thus with increasing disorder, the CO region bounded by CD-1 and CD-2, shrinks forming the profile seen in (b).
 
\begin{figure}
\vspace{.2cm}
\centerline{
\includegraphics[width=9.0cm,height=9.0cm,angle=0,clip=true]{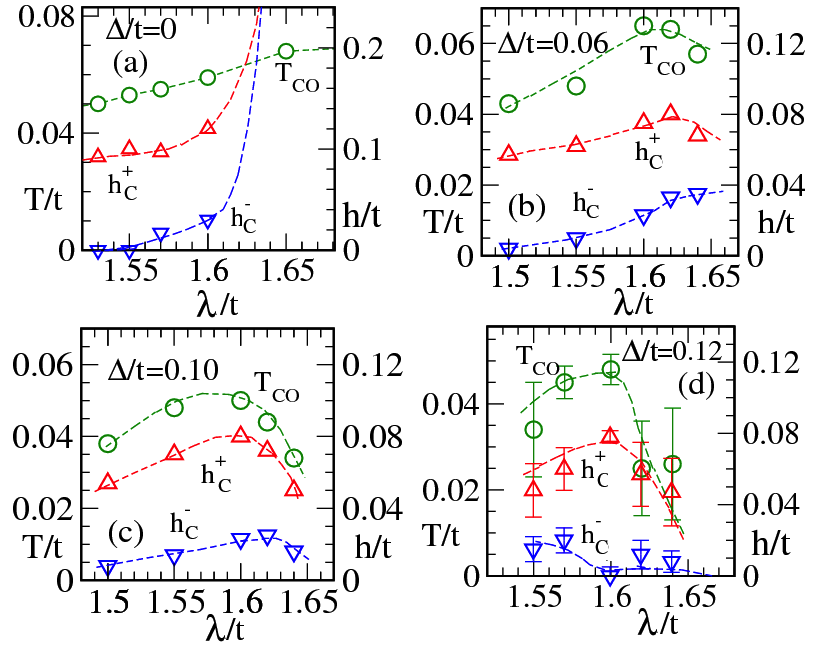} }
\vspace{.2cm}
\caption{Color online: (a) to (d) show the variation of the disorder averaged ${T_{CO}}$, $h^{+}_{c}$ and $h^{-}_{c}$ for $\Delta/t=0, 0.06, 0.1, 0.12$ respectively. 
The $h^{+}_{c}$ and $h^{-}_{c}$ are calculated at $T/t=0.01$. For $\Delta/t=0$, the melting fields diverge beyond a critical $\lambda$ and the $T_{CO}$ grows, in the $\lambda$ window shown. Increasing $\Delta$ not only suppresses the overall melting scales, but also changes the $\lambda$ (or $r_A$) dependence qualitatively. Within the same $\lambda$ window, increasing disorder causes a gradual downturn in all the melting scales at weak disorder and finally strongly suppresses these scales at $\Delta/t\sim 0.12$. Error bars for the largest disorder case (d) are the biggest and are given as an estimate of the maximum error in the numerics.}
\label{f-8}
\end{figure}

 Thus at any finite disorder cross section of the phase diagram, the $T_{CO}$ is  non monotonic in $\lambda$, with the overall magnitude getting suppressed and the peak shifted to lower $\lambda$ compared to the clean case. This is seen in the four panels in Fig.\ref{f-8}, where the $T_{CO}$ are shown in green circles for $\Delta/t=(0, 0.06, 0.1, 0.12)$. Also shown in Fig.\ref{f-8} are the $h^{+}_{c}$ and $h^{-}_{c}$. With increasing disorder the CO state weakens as seen in the suppression on $T_{CO}$ scales in panels (b)-(d). The $h^{\pm}_{c}$  as a function of $\lambda$ follow the behavior of $T_{CO}$ in these cases. The suppression is largest for $\Delta/t=0.12$. For the $\Delta/t=0$ case the magnetic field converts the CE-CO-I state into a FM-CO-I for $\lambda\geq 1.65$ so the CO melting field diverge.
 
In the experiments, for the systems with small disorder, $\sim 10^{-3}A^2$ (e.g. the Ca family), the magnetic melting scales increase with decreasing $r_A$, for systems with moderate disorder, $ \sim 10^{-2}A^2$, (e.g. the Sr family), the melting scales initially increase with decreasing $r_A$ and then are strongly suppressed and eventually go to zero with further decrease in $r_A$. 

In view of these results we conclude that, the Ca family would fall in the low $\Delta/t\le0.05$ category, the Sr family would be placed at $\Delta/t\sim0.12$. The Ba family would be beyond $\Delta/t\sim0.15$.  We have shown in our earlier work that the numerical data is in the correct ballpark of experiments. \cite{short}.

\section{Conclusions}
Using a real space  Monte-Carlo scheme we have shown how the charge  ordered state in the half-doped manganites is affected by disorder and low temperature magnetic field sweeps. 
We explained the counterintuitive bandwidth dependence of the response to the magnetic field of the CE-CO-I state in the Sr based half doped manganites. 
Finally we provided a universal understanding of the field melting problem across all half doped manganites in terms of the zero field 'disorder-bandwidth' phase diagram. 
Our work indicates the following phenomenology for the nature of the CE-CO-I melting transition. Relatively clean samples, at quasi-static sweep rates, will show a continuous transition. Still at low disorder, increasing the sweep rate will lead to step like feature in the magnetization. Finally at large sweep rates the transition will become abrupt. This phenomenology, for intermediate to fast sweep rates, has been seen in experiments\cite{step-5, step-4}. If disorder is sufficiently strong the phase transition will be rounded even for fast seep rates. Experiments studying the interplay of disorder strength and field sweep rates will be extremely interesting in this regard.

From a more general point of view these results have close connections with random field Ising model and melting of vortex lattices that transcends the particular material discussed here.
\vspace{.2cm}

We acknowledge use of the Beowulf cluster at HRI.
\\

\section {Appendix}
In order to study the evolution of the system with applied magnetic field, 
we track various physical quantities in real space and momentum space.
We compute the distribution of lattice distortions,
$P({Q})=\sum_i\delta({Q}-{Q_i} )$, where $Q_i = \vert {\bf Q}_i \vert$; spatial ${Q}_i,{Q}_j$ correlations, $D_Q(\textbf{q})=\sum_{ij} \langle 
{\bf Q}_i {\bf Q}_j \rangle e^{i \textbf{q}.({ \bf r}_i- {\bf r}_j)}$, 
and spin-spin correlations, $S(\textbf{q})=\frac{1}{N^2}\sum_{ij} \langle {\bf S}_i.{\bf S}_j
\rangle e^{i \textbf{q}.( {\bf r} _i- {\bf r}_j)}$. Angular brackets represent a
thermal average. We also compute the volume fraction of the charge ordered region in the lattice from direct spatial snapshots of the charge distribution. To measure the volume fraction, we tag a site with a particular color if the site has $n > 0.5$ and is surrounded by the four nearest neighbor sites with $n < 0.5$ and vice verse (i.e. a site with local anti-ferro-charge correlation is marked with a particular color). Similarly, if the difference between the charge density at a site with its nearest neighbors is less than a threshold, that site is tagged by a different color, \textit{i.e}., the charge uniform regions are marked by this color. For intermediate cases, we use an interpolative color scheme. The volume fraction is necessary for studying inhomogeneous melting
where the momentum space structure factors are not a good measure of the amount of CO in the system. Further, the spatial snapshots of the real space charge density also directly provide visual information on the melting process. While the indicators above measure the spatial correlations and spatial evolution, the metallic or insulating character is tracked via d.c. conductivity\cite{sk-pm-transp}, $\sigma_{dc}$, and the density of states (DOS), $N(\omega)= \langle {1 \over N} \sum_n \delta(\omega-\epsilon_n) \rangle $, where $\epsilon_n$ are the electronic eigenvalues in some MC background and the angular brackets indicate thermal average. We track all the above quantities as a function of temperature and applied magnetic fields for studying the CO melting phenomenon.

\bibliography{ref_new.bib}
\end{document}